# Pressure-Tuned Magnetism and Bandgap Modulation in Layered Fe-Doped CrCl$_3$

Aya Ali[1], Govindaraj Lingannan[2], Lukas Gries[3], Md Ezaz Hasan Khan[2], Anas Abutaha[4], Kei Uemura[5], Masaki Mito[5], Vladislav Borisov[6], Anna Delin[7,8,9], Olle Eriksson[6,10], Rüdiger Klingeler[3], Mahmoud Abdel-Hafiez*[1,11]

[1]Center for Advanced Materials Research, Research Institute of Sciences and Engineering, University of Sharjah, Sharjah 27272, United Arab Emirates
[2]University of Doha for Science and Technology, Doha, Qatar
[3]Kirchhoff Institute of Physics, Heidelberg University, INF 227, D-69120 Heidelberg, Germany
[4]Qatar Environment and Energy Research Institute (QEERI), Hamad Bin Khalifa University, Qatar Foundation, Doha 34110, Qatar
[5]Graduate School of Engineering, Kyushu Institute of Technology, Fukuoka 804-8550, Japan
[6]Department of Physics and Astronomy, Uppsala University, Uppsala, 751 20, Sweden
[7]Department of Applied Physics, KTH Royal Institute of Technology, 516, SE-75121, Stockholm, Sweden
[8]Swedish e-Science Research Center (SeRC), KTH Royal Institute of Technology, SE-10044 Stockholm, Sweden
[9]Wallenberg Initiative Materials Science for Sustainability (WISE), KTH Royal Institute of Technology, SE-10044 Stockholm, Sweden
[10]Wallenberg Initiative Materials Science for Sustainability (WISE), Uppsala University, Uppsala, 751 20, Sweden
[11]Department of Applied Physics and Astronomy, University of Sharjah, P. O. Box 27272 Sharjah, United Arab Emirates

Chromium-based halide magnets, particularly CrCl$_3$, are emerging materials with highly tunable magnetic properties, making them attractive for applications in heterostructures, spintronic devices, and nanoscale sensors. In this study, we explore the structural, magnetic, vibrational and optical band gap properties under varying pressures. By integrating first-principles calculations with experimental techniques, including Raman spectroscopy, photoluminescence (PL), uniaxial pressure studies (thermal expansion), and magnetization measurements, we unveil the intricate pressure-induced transformations in Fe-doped CrCl$_3$, shedding light on its structural, electronic, and magnetic evolution. At ambient pressure, Raman spectra confirm all expected Raman-active modes, which exhibit blue shifts with increasing pressure. The PL measurements demonstrate an optical bandgap of 1.48 eV at ~0.6 GPa, with a progressive increase in the bandgap under pressure, transitioning slower above 6 GPa due to an isostructural phase transition. Magnetization results under pressure shows two competing magnetic components (FM and AFM) at ambient conditions, where at the lowest temperature and applied field, the FM component dominates. The presence of competing FM and AFM energy scales is confirmed by Grüneisen analysis of the thermal expansion and their uniaxial pressure dependence is determined. The experimental findings agree with theoretical results based on Density functional theory (DFT). In the experiments, we observe a pressure-enhanced ferromagnetic interlayer coupling that is followed by the stabilization of antiferromagnetic ordering, due to weakened direct interlayer interactions. Above 1.2 GPa the FM component of the magnetism is gone in the experimental observations, which is also in good agreement with DFT based theory. The findings reported here underscore the potential of CrCl$_3$ for use in pressure-tunable magnetic and optoelectronic applications, where, e.g., the delicate balance between FM and AFM configurations could have potential for sensor applications. These results provide valuable insights into the pressure-dependent properties of CrCl$_3$, paving the way for its application in pressure-tunable magnetic and electronic devices.

2## I. INTRODUCTION

Significant advancements in two-dimensional (2D) materials research were catalyzed by the groundbreaking isolation of monolayer graphene from graphite by A. Geim and K. Novoselov in 2004 [1,2]. This milestone expanded the horizons of condensed matter physics, leading to the discovery of diverse and novel properties in a wide range of 2D quantum materials. Among these, van der Waals (vdW) materials have garnered particular interest due to their unique magnetic, electrical, and optoelectronic properties, making them highly promising for next-generation technologies [3–7]. These materials are defined by weak interlayer van der Waals forces and strong intralayer covalent bonding, giving rise to remarkable magnetic behaviors with significant anisotropy [8–10]. Furthermore, the investigation of spin dynamics and the establishment of long-range ferromagnetic (FM) order in 2D vdW systems underline their substantial potential for spintronic applications, where control over spin degrees of freedom can revolutionize information processing and storage technologies [11,12]. An important class of vdW 2D materials is the family of transition metal trihalides (TMTHs), which have the general formula $MX_3$, where M represents a transition metal cation, and X is a halogen anion. TMTHs have emerged as pivotal systems for exploring monolayer magnetism, particularly after the experimental discovery of stable out-of-plane ferromagnetic spins in chromium triiodide ($CrI_3$) in 2017 [2,13,14].

The magnetic properties of TMTHs are rooted in the electronic configuration of the transition metal atoms, specifically the partially filled *d*-orbitals. Electron-electron Coulomb repulsions within these orbitals generate strong magnetic exchange interactions, which are key to the emergence of magnetic order. In addition, spin-orbit coupling and crystal field effects contribute to the magnetic anisotropy observed in these materials, enabling the stabilization of out-of-plane spin orientations. The interplay of strong exchange interactions and anisotropy helps overcome these fluctuations, stabilizing long-range magnetic order even in monolayer systems [15–17]. This unique combination of electronic and magnetic features positions TMTHs as ideal candidates for advancing fundamental research and practical applications in spintronics and quantum information science.

Chromium-based halide magnets, specifically $CrX_3$ (X = I, Br, Cl), exhibit highly tunable magnetic properties. Bulk $CrI_3$ shows FM behavior with a transition temperature ($T_C$) of 61 K [18-22]. $CrBr_3$ exhibits FM behavior at approximately 34 K [23,24], while $CrCl_3$ demonstrates AFM behavior at 14 K, with in-plane FM coupling and interlayer AFM interactions [25,26]. The magnetic transition temperature and unit cell volume increase with the substitution of Cl by Br or I. This trend occurs because the larger halogen ions increase the Cr–Cr distance, weakening the AFM coupling driven by direct exchange interactions. Simultaneously, the larger halogens enhance spin-orbit coupling, which contributes to the magnetic anisotropy of these materials [27]. As direct exchange weakens, superexchange interactions become more significant, lowering the system's total energy via the virtual hopping of electrons between neighboring Cr ions through a halogen ion. According to the Goodenough-Kanamori-Anderson (GKA) rules, the Cr–X–Cr bond angle determines the type of superexchange interaction: FM coupling is favored when the angle is near 90°, while AFM coupling occurs at angles close to 180° [29, 30]. In fact, first-principles calculations predicts that significant Cr-Cr magnetic pair interactions change sign from ferromagnetic (FM) to antiferromagnetic (AFM) in monolayer $CrI_3$ as the Cr-Cr distance is altered [31].

Hydrostatic pressure is an effective and impurity-free tool for tuning the properties of vdW-layered 2D materials. It allows precise modulation of bond lengths, bond angles, and interatomic distances, leading to controlled tuning of the crystal structure [31,32]. In $CrI_3$, $T_C$ was increased from 61.5 K to 66 K under 3 GPa of pressure. However, further increasing the pressure reduced $T_C$, reaching 10 K at 21.2 GPa. Theoretical calculations suggest that the initial Cr–I–Cr bond angle of 95° decreases to 90° at the point of maximum $T_C$. As pressure increases beyond this, the angle decreases further to 85° at 25 GPa. Additionally, a semiconductor-to-metal transition was observed above 22 GPa. Raman measurements revealed distinct changes in behavior at 4.4 GPa: below this pressure, only minor variations were observed in Raman modes, but above 4.4 GPa, the rate of change accelerated, with two Raman modes disappearing at 7.2 GPa. All Raman modes vanished above 17.1 GPa, suggesting sample metallization and pressure-induced lattice deformation (isostructural transition, IST) [31,33]. In $CrBr_3$, high-pressure Raman measurements revealed the emergence of a new peak at 2.5 GPa and a change in the compression rate of the lattice parameter *c* near 7 GPa, possibly indicating an IST. Additionally, the $E_g^1$ and $A_g^1$ Raman modes merged at 9.5 GPa. The $T_C$ of $CrBr_3$ decreases under pressure at a rate of $dT_C/dP$ = -4.1 K/GPa, with complete suppression of FM behavior expected around 8.4 GPa. Another anomaly in the $A_g^1$ mode was observed at 15 GPa, potentially related to metallisation as in its sister compound $CrI_3$. The Cr–Br–Cr bond angle decreases from 95° at 0 GPa to 92.5° at 10 GPa. This reduction in $T_C$ with increasing pressure is attributed to the decreased magnetocrystalline anisotropy energy and reduced in-plane magnetic coupling between Cr ions [34–37].

In $CrCl_3$, an IST and the disappearance of the $A_g^3$ Raman mode were observed, along with a maximum indirect bandgap of 1.8 eV at 11 GPa. Pressure-induced magnetic phase transitions from FM to AFM were also reported likely driven by strong coupling between the spin and lattice degrees of freedom. An electronic topological transition (ETT) was observed at ~30 GPa [26,38,39]. $CrCl_3$ exhibits short-range FM order at 17 K and long-range AFM order at 14 K [27]. Moreover, it shows a favorable magnetic entropy change and higher relative cooling power compared to $CrI_3$ [39]. With in-plane magnetic anisotropy and A-type AFM behavior, $CrCl_3$ is a promising candidate for developing advanced spintronic devices by exploring magnetic quantum phase transitions in detail [25]. Doping $CrI_3$ with small amounts of Mn or V shifts $T_C$ to lower temperatures. Mn-doped $CrI_3$ shows higher saturation magnetization, while V-doped $CrI_3$ exhibits higher coercivity and increased magnetocrystalline anisotropy. Theoretical predictions suggest the presence of direct and narrow bandgaps, making these doped samples suitable candidates for future spintronic applications [40]. Similarly, alkali metal-doped monolayer $CrCl_3$ shows an enhancement of $T_C$ from 23 K to 66 K. Theoretical calculations predict a transition from a semiconducting to a Dirac half-metallic phase, as well as the quantum anomalous Hall effect in Na- and K-doped samples [41,42,43].

Motivated by these findings, we investigate the effects of chemical and physical pressure on the $CrCl_3$ system. This study focuses on the structural, magnetic, and optical bandgap properties of Fe-doped $CrCl_3$ (Fe = 0.1, 0.2, 0.3, 0.4, 0.5) with experimental probes that involve ambient and high-pressure Raman spectroscopy and photoluminescence (PL) measurements, dilatometry studies, and magnetization studies under varying temperature, magnetic field, and pressure conditions. The experimental investigations are complemented with first-principles theory based o density functional theory.

## II. EXPERIMENTAL AND COMPUTATIONAL DETAILS

Chromium trichloride ($CrCl_3$) was synthesized using the chemical vapor transport (CVT) method. High-purity $CrCl_3$ powder is sealed in an evacuated quartz ampoule along with a transport agent, iodine, which facilitates the sublimation and recrystallization of $CrCl_3$ in a temperature gradient. The ampoule is placed in a two-zone furnace, where the source zone is maintained at a higher temperature (650–700°C) and the growth zone at a lower temperature (550–600°C) [29]. Over 7-days, $CrCl_3$ vapor gradually deposits in the cooler zone, forming well-defined layered single crystals. To introduce iron (Fe) into the $CrCl_3$ lattice, a controlled amount of $FeCl_3$ and additional Fe precursors are added to the $CrCl_3$ starting material, ensuring a stoichiometric composition corresponding to $Fe_xCr_{1-x}Cl_3$ with Fe concentrations $x$ = 0.1, 0.2, 0.3, 0.4, and 0.5. The mixture is then sealed in an evacuated quartz ampoule with iodine as the transport agent and subjected to the same CVT growth conditions as pure $CrCl_3$. Ambient and high-pressure micro-Raman scattering measurements were performed using a Renishaw micro-Raman spectrometer. A 532 nm (2.33 eV) diode laser was employed as the excitation source, coupled with a 2400 lines/mm grating, enabling a diffraction-limited laser spot with a diameter of approximately 1.0 μm and a spectral resolution of around 1 cm$^{-1}$. A 10 X long-working distance objective lens (numerical aperture = 0.50) was used to focus the laser beam onto the sample surface. For photoluminescence (PL) measurements, an 1800 lines/mm grating was utilized. In situ, high-pressure Raman and PL spectra were obtained using a symmetric diamond anvil cell (DAC) with diamond anvils of 500 μm culet size. A 300 μm thick SS301 stainless steel gasket was pre-indented to a thickness of 100 μm to enhance its hardness, and a 200 μm diameter hole was drilled in the center of the gasket using a tungsten carbide (WC) drill bit to create the sample chamber. Silicon oil was used as the pressure-transmitting medium. The pressure inside the sample chamber was measured using the ruby fluorescence method, with ~5 μm ruby chips placed alongside the sample for calibration. During high-pressure measurements, a waiting period of at least 30 minutes was allowed between each pressure increment to ensure that thermal and pressure equilibria were reached. The Raman spectra were fitted with Lorentzian profiles to extract mode frequencies and peak widths. The ac magnetization was measured in a Quantum Design superconducting quantum interference device (SQUID) magnetometer by using a miniature DAC. The amplitude and frequency of ac field were 3.8 Oe and 10 Hz, respectively. The culet size of diamond anvils was 0.8 mm. The sample hole with the diameter of 0.3 or 0.4 mm was prepared using an electric discharge machine in a CuBe gasket with initial thickness of 0.3 mm. Several pieces of crystals and liquid-like pressure transmitting medium Apiezon J oil (M&I Materials Limited) were held into the sample cavity with some ruby balls. The pressure at room temperature was evaluated by measuring the fluorescence of ruby balls inside the sample chamber. The increase in pressure at liquid helium temperature against that at room temperature is at most 10%. Measurements were performed on single-crystalline flakes of $CrCl_3$ and $Cr_{0.5}Fe_{0.5}Cl_3$. Magnetisation measurements were performed using Quantum Design's Magnetic Properties Measurement System (MPMS3) in the temperature range of 2 K to 300 K and in fields up to 7 T applied along the crystallographic $c$ direction. High-resolution dilatometry measurements were performed in a three-terminal capacitance dilatometer from Küchler Innovative Measurement Technologies [54,55]. The measurements have been done in a home-built setup in a Variable Temperature Insert (VTI) of an Oxford Instrument magnet system at zero field and B = 15 T applied along the direction of measured length changes [56]. The sample thickness in the measurement direction, i.e., the crystallographic c axis, was 0.149 mm. The linear thermal expansion coefficients $\alpha_c = 1/L_c(300K) \times \partial L_c(T)/\partial T$ are derived from the relative length changes. To explore pressure-dependent properties of $CrCl_3$ from first principles we used density functional theory (DFT) with static mean-field correlation available in the VASP code [53]. The exchange-correlation energy was described using the PBE-parametrized generalized-gradient approximation. The rotationally invariant Lichtenstein formulation of GGA+$U$ with $U$ = 3 eV and $J_H$ = 0.6 eV was used for a more accurate description of electronic correlations, needed for correctly capturing the insulating character of $CrCl_3$. The chosen values of $U$ and $J_H$ parameters provide a reasonable value of the electronic band gap and the Cr magnetic moment magnitude and are kept fixed for all studied pressures.

## III. RESULTS

### A. RAMAN MEASUREMENTS

Raman spectroscopy is a powerful tool for investigating magnetic materials, providing insights into their magnetic excitations, such as energy, symmetry, and statistical properties. It also allows the study of interactions between magnetism and other physical phenomena, including structural phase transitions and spin-phonon coupling [47]. By examining how magnetic order influences phonon behavior, Raman spectroscopy becomes instrumental in understanding both the magnetic and lattice dynamics of materials. The motivation for studying the effects of chemical doping and hydrostatic pressure on $CrCl_3$ lies in the need to understand how these external perturbations influence its structural, vibrational, and magnetic properties. $CrCl_3$, a prominent member of the transition metal trihalides family, has attracted significant attention due to its layered van der Waals structure and intriguing magnetic behavior. Chemical doping with Fe introduces controlled variations in the electronic structure and magnetic interactions by substituting Cr sites with Fe atoms. This substitution alters the lattice dynamics, modifies the vibrational modes, and impacts the overall stability and symmetry of the crystal.

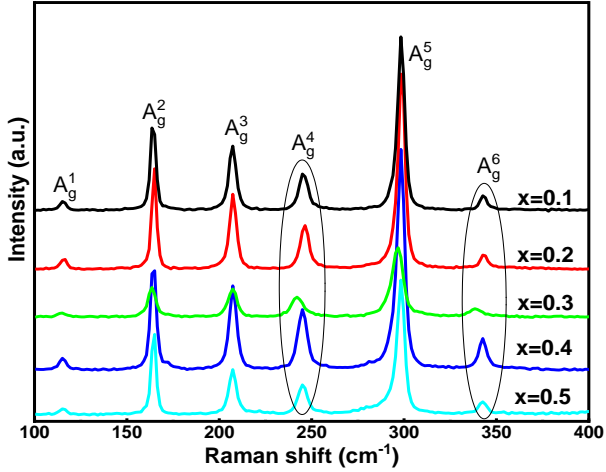

Figure 1: Measured Raman spectra of $Fe_xCrCl_3$ (x = 0.1, 0.2, 0.3, 0.4, and 0.5) samples under ambient conditions using a 532 nm laser source.

By systematically varying the Fe content (x = 0.1, 0.2, 0.3, 0.4, and 0.5), we aim to uncover the role of dopant concentration in tuning the phononic and electronic properties of $CrCl_3$.

On the other hand, hydrostatic pressure provides a complementary approach by compressing the lattice uniformly, thereby altering interlayer and intralayer bonding interactions without introducing chemical impurities. Studying $CrCl_3$ under high pressures (up to 16 GPa) allows us to probe the pressure-induced modifications in vibrational modes, bandgap energies, and magnetic ordering. Such studies are crucial for identifying phase transitions, pressure-induced anisotropy, and changes in magnetic coupling. Figure 1 displays the Raman spectra of the Fe-doped $CrCl_3$ samples with varying Fe content (x = 0.1, 0.2, 0.3, 0.4, and 0.5). The experimental Raman spectra for these samples are in excellent agreement with the theoretically predicted spectra of pristine $CrCl_3$ crystals, indicating high crystalline purity of the synthesized materials [38,48,49]. For undoped $CrCl_3$, the observed Raman active modes appear at 116.5 cm$^{-1}$, 165.3 cm$^{-1}$, 208.6 cm$^{-1}$, 246.5 cm$^{-1}$, 300.0 cm$^{-1}$, and 344.5 cm$^{-1}$ because of the intralayer vibration. These modes correspond well to calculated Raman peaks at 118 cm$^{-1}$, 166 cm$^{-1}$, 207 cm$^{-1}$, 250 cm$^{-1}$, 300 cm$^{-1}$, and 345 cm$^{-1}$, which are assigned as $A_g^1$(Cl), $A_g^2$(Cl), $A_g^3$, $A_g^4$, $A_g^5$, $A_g^6$(Cr), phonon modes, respectively [38,48,49]. These modes reflect the translational motion (T') of Cr atoms in the $CrCl_3$ structure.

The parent compound $CrCl_3$ shares the same crystal symmetry (C2/m) as $CrI_3$ under ambient conditions, resulting in similar Raman vibrational phonon modes at room temperature [33,38,50,51]. The sister compound $CrBr_3$, on the other hand, crystallizes in a rhombohedral structure with $\bar{R3}$ symmetry at room temperature. It exhibits four $E_g$ and two $A_g$ modes, demonstrating symmetry-dependent differences between these compounds [34]. $FeCl_3$ exhibits $C_{2h}^3$ symmetry, with three $A_g$ and three $E_g$ vibrational modes [49,52]. Upon doping $CrCl_3$ with Fe (x = 0.1 to 0.5), no significant shifts in the Raman peaks were detected suggesting that Fe incorporation did not cause drastic changes in the crystal structure. The measured Raman peak positions for the Fe-doped samples are summarized in Table 1.

Table 1: Measured Raman Modes of Vibrations for $Fe_xCrCl_3$ (x = 0.1, 0.2, 0.3, 0.4, and 0.5)

| x | $A_g^1$ (cm$^{-1}$) | $A_g^2$ (cm$^{-1}$) | $A_g^3$ (cm$^{-1}$) | $A_g^4$ (cm$^{-1}$) | $A_g^5$ (cm$^{-1}$) | $A_g^6$ (cm$^{-1}$) |
|---|---|---|---|---|---|---|
| 0.1 | 115.3 | 164.2 | 207.1 | 245.6 | 298.3 | 343.0 |
| 0.2 | 115.7 | 164.7 | 207.4 | 246.2 | 298.5 | 343.2 |
| 0.3 | 114.7 | 163.5 | 207.1 | 242.2 | 296.3 | 338.7 |
| 0.4 | 115.2 | 164.3 | 207.4 | 245.0 | 298.1 | 342.6 |
| 0.5 | 115.3 | 164.5 | 207.2 | 245.1 | 298.1 | 342.5 |

The absence of significant Raman peak shifts with increasing Fe doping suggests that the crystal symmetry and lattice dynamics remain relatively unchanged across the doping range. Minor variations in peak intensities may arise from slight differences in sample quality, as seen for Fe = 0.2 and 0.3. These results further validate that Fe doping does not disturb the fundamental structural framework of $CrCl_3$, making the Fe-doped $CrCl_3$ system a promising candidate for further exploration in spintronic applications.

### B. Pressure dependent of Raman study

High-pressure Raman spectroscopy was employed to investigate the vibrational transitions in $CrCl_3$, the sample with the highest Fe concentration in this series. Measurements were conducted at pressures up to approximately 16 GPa to explore the evolution of phonon modes under compression and identify potential pressure-induced phase transitions. At ambient pressure (0 GPa), the Raman spectra display all six expected Raman-active modes [38,48,49]. As pressure increases, all modes exhibit a smooth shift towards higher frequencies (blue shift), with their behavior illustrated in Figure 2(a). The pressure-induced shifts with the rate of 4.85 cm$^{-1}$/GPa (Ag1), 2.17 cm$^{-1}$/GPa (Ag2), 1.18 cm$^{-1}$/GPa (Ag4), 3.10 cm$^{-1}$/GPa (Ag5) and 4.85 cm$^{-1}$/GPa (Ag6).

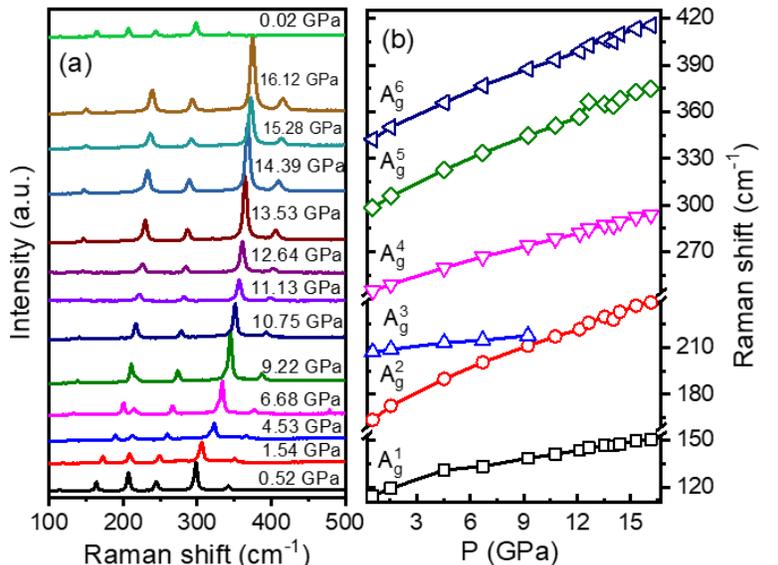

Figure 2: (a) Measured Raman spectra for Fe-doped $CrCl_3$ (x=0.5) at selected pressures at room temperature, and (b) pressure dependence of the observed Raman modes.

Notably, the $A_g^3$ mode initially shifts at 4.65 cm$^{-1}$/GPa but disappears completely around 10.75 GPa, signaling an IST

[5,26,38]. This behavior mirrors the phonon dynamics observed in parent $CrCl_3$, where the $A_g^3$ mode vanishes near 9.9 GPa, as well as in $CrI_3$ and $CrBr_3$, which exhibit the disappearance of similar modes at 7.2 GPa and 9.5 GPa, respectively [33,35]. The variation in the critical pressure across these compounds suggests that Fe doping significantly influences the phase stability and phonon dynamics of $CrCl_3$.

A comparative analysis with undoped $CrCl_3$ reveals differences in phonon shift rates: 3.06 cm$^{-1}$/GPa ($A_g^1$), 5.88 cm$^{-1}$/GPa ($A_g^2$), 1.52 cm$^{-1}$/GPa ($A_g^3$), 3.72 cm$^{-1}$/GPa ($A_g^4$), 6.18 cm$^{-1}$/GPa ($A_g^5$), and 5.73 cm$^{-1}$/GPa ($A_g^6$) [35]. These discrepancies suggest that Fe doping alters the local crystal environment, modifying the interatomic forces and lattice dynamics, which results in distinct phonon behavior under pressure. Upon decompression, all vibrational modes reappear, exhibiting redshifts, confirming that the structural transition is reversible with no permanent alteration of the crystal structure. The reversible nature of these transitions highlights the robustness and stability of the $CrCl_3$ lattice under high-pressure conditions. This study provides valuable insights into how Fe doping modulates the vibrational properties and structural stability of $CrCl_3$, shedding light on the interplay between chemical doping, pressure, and phonon dynamics in magnetic materials [33,35,38,49].

### C. Pressure-Dependent Photoluminescence

Figure 3(a) shows the high-pressure photoluminescence (PL) spectra of Fe-doped $CrCl_3$ (x=0.5), while Figure 3(b) illustrates the evolution of the optical energy gap as a function of pressure up to 14.41 GPa. The PL spectrum at 0.65 GPa reveals an optical energy gap of 1.48 eV, which closely corresponds to its indirect bandgap of parent $CrCl_3$ (1.67 eV) [26]. As the pressure increases, the broad PL peak shifts towards shorter wavelengths, indicating a pressure-induced increase in the energy bandgap. However, above 6 GPa, the rate of increase in the bandgap slows down, which may be attributed to the onset of an IST.

For the parent $CrCl_3$, a reversal in the bandgap trend has been reported at 10.2 GPa, where an IST occurs, accompanied by a magnetic phase transition from ferromagnetism (FM) to antiferromagnetism (AFM). However, in the synthesized Fe-doped $Cr_{1-x}Fe_xCl_3$ (x=0.5), no such switching behavior is observed in the energy bandgap trend under pressure. The absence of FM-to-AFM switching suggests that Fe doping plays a critical role in stabilizing the magnetic state under high pressure. Since Fe exhibits ferromagnetic behavior, it may prevent the magnetic transition seen in undoped $CrCl_3$, contributing to the lack of switching in the energy bandgap [26]. Upon decompression, the system returns to its original phase, with the energy gap recovering to 1.48 eV at 0.59 GPa. This reversibility indicates that the structural and electronic changes induced by pressure are not permanent, highlighting the resilience of the $CrCl_3$ lattice.

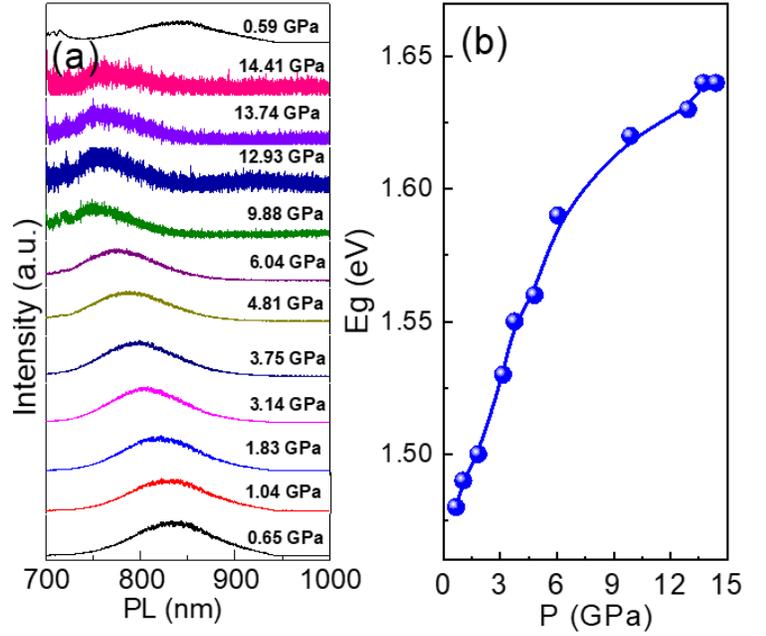

Figure 3: (a) Measured PL spectra of Fe-doped $CrCl_3$(x=0.5) at selected pressures at room temperature, and (b) calculated energy band gap as a function of pressure from the PL study.

### D. Pressure dependent Magnetization

The ferromagnetic nature of $CrCl_3$ studied at hand is illustrated by a sharp increase of the magnetization, at low temperatures, while long-range antiferromagnetic order is signalled by a peak in M (T,B = 10 mT) at the Néel temperature ($T_N$) = 13.5 K. The saturation magnetization, at T = 1.8 K, assumes $M_{sat}$ = 2.88(3) $\mu_B$/f.u. which agrees with 2.9 $\mu_B$/f.u. reported in Ref. [39] but is slightly smaller than 3.0 $\mu_B$/f.u. from Ref. [27]. Figure 4 shows Temperature dependence of the magnetization measured on $Cr_{0.6}Fe_{0.4}Cl_3$ at B = 0.01T and B = 1T applied along the crystallographic c axis. In addition, Fisher's specific heat reveals a distinct anomaly at $T_S$ = 251(1) K associated with the structural phase transition reported previously [27].

Figure 5(a) shows the magnetization data of $CrCl_3$ as a function of temperature under various fixed pressures up to 2.5 GPa.

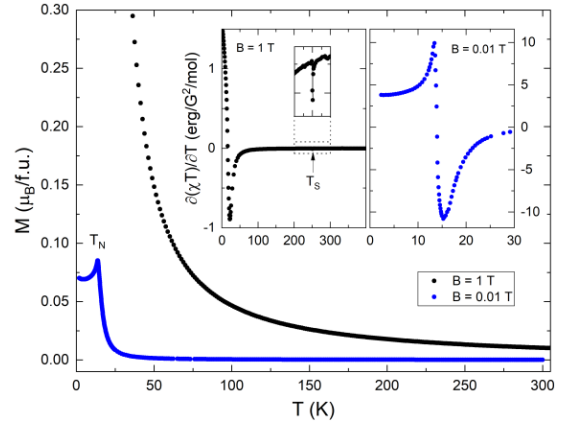

Figure 4. Temperature dependence of the magnetization measured on $Cr_{0.5}Fe_{0.5}Cl_3$ at B = 0.01T and B = 1T applied along the crystallographic c axis. Insets: Fisher's specific heat $\partial(\chi T)/\partial T$ and highlight the anomaly at the structural transition at $T_S$ = 251 K (left) and $T_C$ (right).

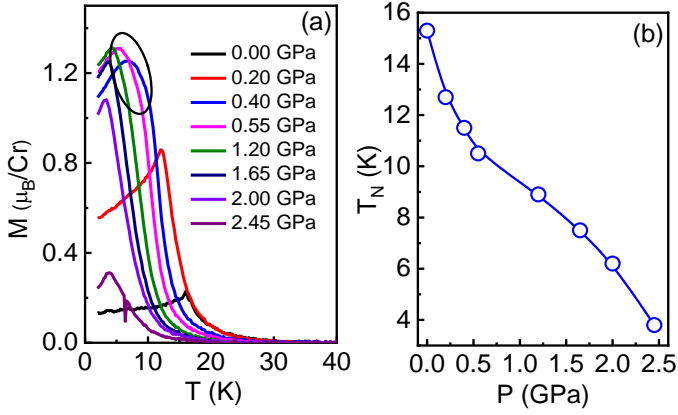

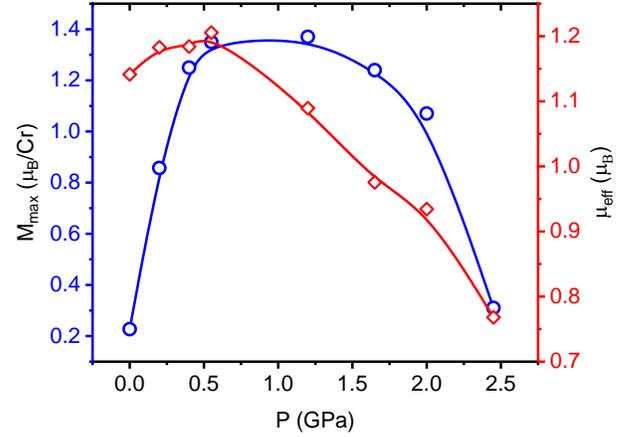

Figure 5: (a) Measured temperature-dependent magnetization data under various fixed pressures up to 2.5 GPa at 10 Oe, from 40 K down to 2 K in the field cooling mode, and (b) Néel temperature ($T_N$) as a function of pressure up to 2.5 GPa.

Figure 6: Pressure dependence of the maximum magnetic moment ($M_{max}$) at the transition temperature ($T_N$) (blue open symbols) and the effective paramagnetic moment ($\mu_{eff}$) (red diamond symbols).

With increasing pressure, the M(T) profile shifts towards lower temperatures, and the magnetic moment coefficient increases rapidly up to 0.4 GPa. However, the rate of increase slows down beyond this pressure, reaching a maximum of 1.2 GPa, followed by a sharp decrease above 1.65 GPa. Concurrently, the sharp cusp associated with the antiferromagnetic transition broadens under pressures of 0.4, 0.55, and 1.2 GPa, indicating the emergence of a small ferromagnetic-like behavior, as highlighted by the black circle in the figure. We have also calculated $T_N$ as a function of pressure using the Curie-Weiss law, as shown in Figure 5(b).

The rate of decrease in $T_N$ is particularly high in the low and high-pressure regions, i.e., below 0.5 GPa and above 1.5 GPa. The $T_N$ decrease confirms that the pressure reduces the Cr-Cr exchange by altering the interlayer distance and bond angle [31,33]. At 3.2 GPa, a magnetic anomaly was observed, whereas at 4.4 GPa, the magnitude is less than that of the scatter in data of the measurements. Additionally, we analyzed the maximum magnetic moment ($M_{max}$) at the transition temperature ($T_N$) and the effective paramagnetic moment ($\mu_{eff}$) for each pressure, as shown in Figure 6 (blue open circles and red open diamonds, respectively).

From these plots, it is evident that the $M_{max}$ increases up to 1.2 GPa, coinciding with the emergence of ferromagnetic-like behavior. However, $T_N$ decreases monotonically with pressure, while the positive Weiss temperatures indicate that intraplanar ferromagnetic interactions dominate the magnetic behavior [27].

The increase in magnetic moment up to 1.2 GPa suggests enhanced interlayer coupling, which favors a ferromagnetic-like interaction. Beyond 1.2 GPa, the weakening of direct interlayer interactions leads to the stabilization of antiferromagnetic (AFM) ordering in the Cr-Cr layers, possibly due to pressure-induced changes in bond angles [33].

The monotonic decrease in $T_N$ with increasing pressure also suggests a deviation in the bond angles, further supporting the AFM order. The initial enhancement in magnetization at low pressures reflects the ferromagnetic-like analogy, which reverts to AFM order with increasing pressure [27]. This is corroborated by the effective paramagnetic moment ($\mu_{eff}$), which initially increases up to 0.55 GPa before decreasing to 2.45 GPa. The $\mu_{eff}$ values remain significantly lower than the expected value for $Cr^{3+}$ (3.87 $\mu_B$), highlighting the complex magnetic interactions in the $CrCl_3$ system.

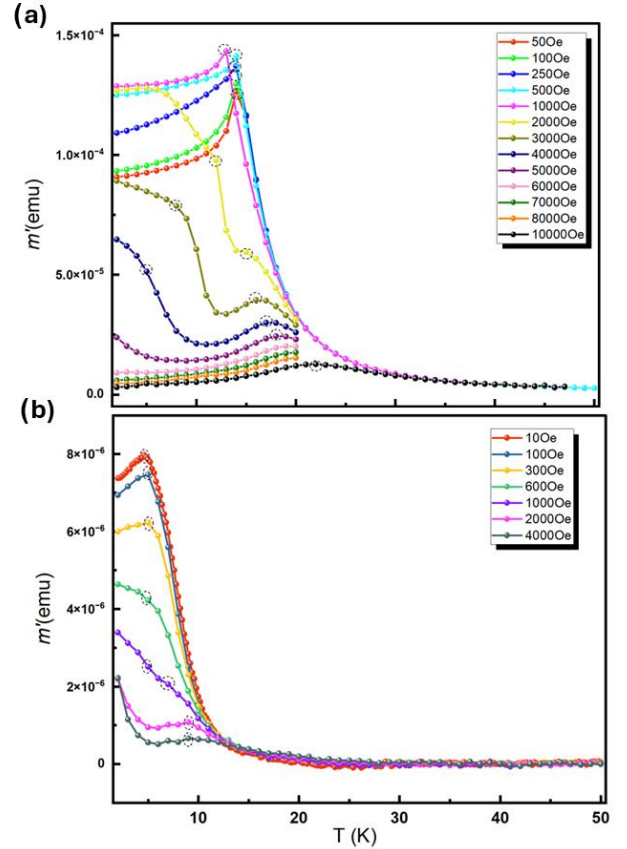

Figure 7: Temperature dependence of magnetization under a fixed pressure of 0 GPa (left) and 1.2 GPa (right): (left) applied magnetic fields ranging from 50 Oe to 1 kOe, and (right) applied magnetic fields from 10 Oe to 4 kOe.

To further investigate the effect of magnetic field under pressure, M(T) measurements were performed at 0 GPa and 1.2 GPa under varying magnetic fields (Fig. 7), ranging from 50 Oe to 10 kOe, as shown in Figure 7. At 50 Oe, an antiferromagnetic transition is observed at 14.9 K. As the magnetic field increases to 1 kOe, the transition shifts to lower temperatures, and the magnetic moment increases [25]. When the applied magnetic field is increased below 3 kOe, the parent $CrCl_3$ exhibits a similar behavior. In this case, the magnetic moments are aligned ferromagnetically within layers, and the layers are stacked antiferromagnetically above, which exhibits ferromagnetic-like behavior because all of the magnetic moments are aligned parallel to the applied magnetic field [25,27].

Further, increasing the applied magnetic field above 1 kOe, our samples cusp in the transition region also diminishes, and ferromagnetic-like behavior becomes more pronounced, similar to what was observed with increasing pressure up to 1.2 GPa. At higher fields, above 1,000 Oe, the ferromagnetic nature becomes more prominent, likely due to the alignment of crystallites along the field direction.

### E. Uniaxial Pressure Effects: Thermal Expansion and Grüneisen Scaling

Our studies of the thermal expansion along the $c$ axis provides further information on the magneto-elastic coupling in $CrCl_3$ and on the uniaxial pressure dependence of $T_N$. As it is displayed in Fig. 8a, the $c$ axis expands quasi-linearly upon heating above ∼80 K. At high temperature, this changes only marginally when applying a high magnetic field of 15 T. Upon cooling, anomalies in the vicinity of $T_N$ and clear effects of the magnetic field on the relative length changes signal significant magneto-elastic coupling. This is particularly evident if the thermal expansion coefficient $\alpha_c$ is considered (Fig. 8b). Specifically, upon cooling it displays a jump-like decrease at $T_N$ and a change in slope at $T^*$ ∼ 17 K. When applying $B$ = 15 T, the jump at $T_N$ vanishes and $\alpha_c$ decreases in the vicinity of $T_C$, i.e., up to about 30 K, but increases at higher temperatures. It is straightforward to attribute this behavior to field-induced suppression of the antiferromagnetically ordered phase [27] and the increase of ferromagnetic correlations above $T_N$, i.e., shift of associated ferromagnetic entropy to higher temperatures. This behavior in particular shows that short-range ferromagnetic correlations are associated with an increase in $\alpha_c$. Hence, according to the Grüneisen relation $\partial\ln(\epsilon)/\partial p_i = V_m \alpha_i/c_p$, our data imply the positive uniaxial pressure dependence of the energy scale $\epsilon_1$ driving in-plane ferromagnetic correlations [57,58]. In contrast, the jump-like decrease in $\alpha_c$, at $T_N$, unequivocally demonstrates that $\partial\epsilon_2/\partial p_c$ is negative, with $\epsilon_2$ being the driving energy of long-range antiferromagnetic order.

This is further illustrated in Fig. 9 where $\alpha_c$ is compared to recently reported specific heat data [27]. As shown in Fig. 9, the jump in $\alpha_c$ at $T_N$ corresponds to the sharp anomaly in $c_p$. In addition, the regime of particularly pronounced anomalous entropy changes between $T_N$ and $T^*$ corresponds to a regime of decreasing $\alpha_c$. The comparison of $c_p$ and $\alpha_c$ is illustrated by the Grüneisen ratio $\gamma_c = \alpha_c/c_p$ in Fig. 9c. Here, we plot the total Grüneisen scaling.

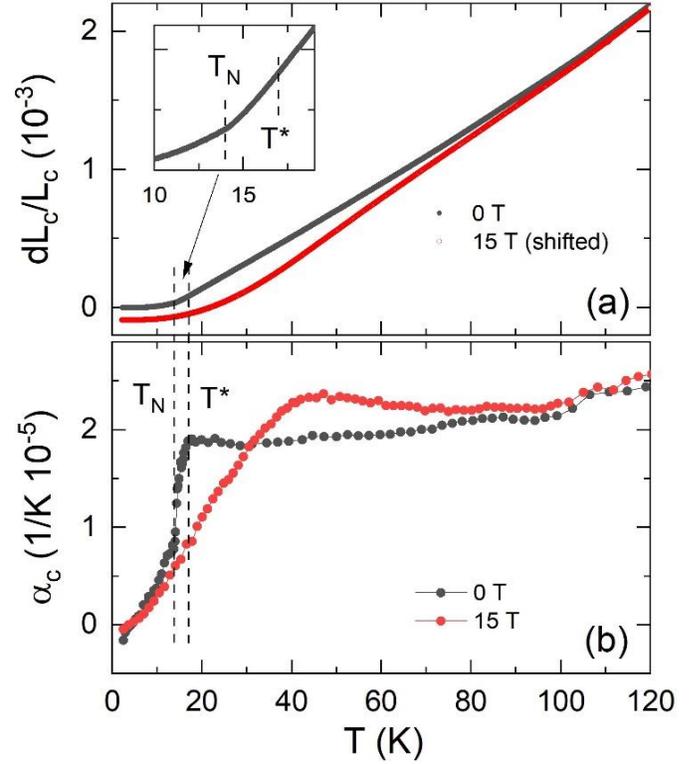

Figure. 8. (a) Uniaxial length changes and (b) thermal expansion coefficient αc of $CrCl_3$. Vertical dashed lines are guide to the eye. Data at 15 T have been shifted by −7 × 10−5 to match the zero field data at 120 K and to illustrate the field effect.

The magnetic Grüneisen ratio which in principle can be calculated by subtracting the phonon contributions to $c_p$ and $\alpha_c$ cannot be obtained reliably for the data at hand. As illustrated in Fig. 9a, we interpret the as magnetic reported specific heat in Ref. [27] (not shown here) as only a very small portion of the total heat capacity which in the temperature regime under study is by far dominated by magnetic degrees of freedom. The background estimated in Ref. [59] as well as our own studies on isostructural materials implies that the specific heat, at $T_N$, is by far (≫ 90 %) dominated by magnetic entropy changes. This analogously holds for the thermal expansion where the phonon background can be estimated from $c_p^{bgr}$ via phononic Grüneisen scaling. The total Grüneisen scaling hence provides meaningful information on magnetic degrees of freedom in $CrCl_3$.

As seen in Fig. 9 our data imply three distinct temperature regimes: (1) Above $T^*$, $\gamma_c$ increases upon cooling and assumes a broad maximum around $T^*$. (2) At $T_N \leq T \leq T^*$, $\gamma_c$ decreases which coincides with the decrease of $\alpha_c$ in this temperature regime. This behaviour corresponds to the increase in $c_p$ which signals the significance of in-plane ferromagnetic short-range correlations [27,39]. (3) At $T_N$, we observe is a negative cusp in $\gamma_c$ which is associated with a jump-like decrease in $\alpha_c$ at the antiferromagnetic ordering transition. The fact that $\alpha_c$ decreases at $T_N$ as well as in temperature regime (2) suggests that the antiferromagnetic interactions driving long-range order are relevant in regime (2), too.

Qualitatively, as mentioned above, the sign of the anomaly at $T_N$ implies $\partial\epsilon_2/\partial p_c < 0$ and it is straightforward to identify $\epsilon_2$ with $T_N$.

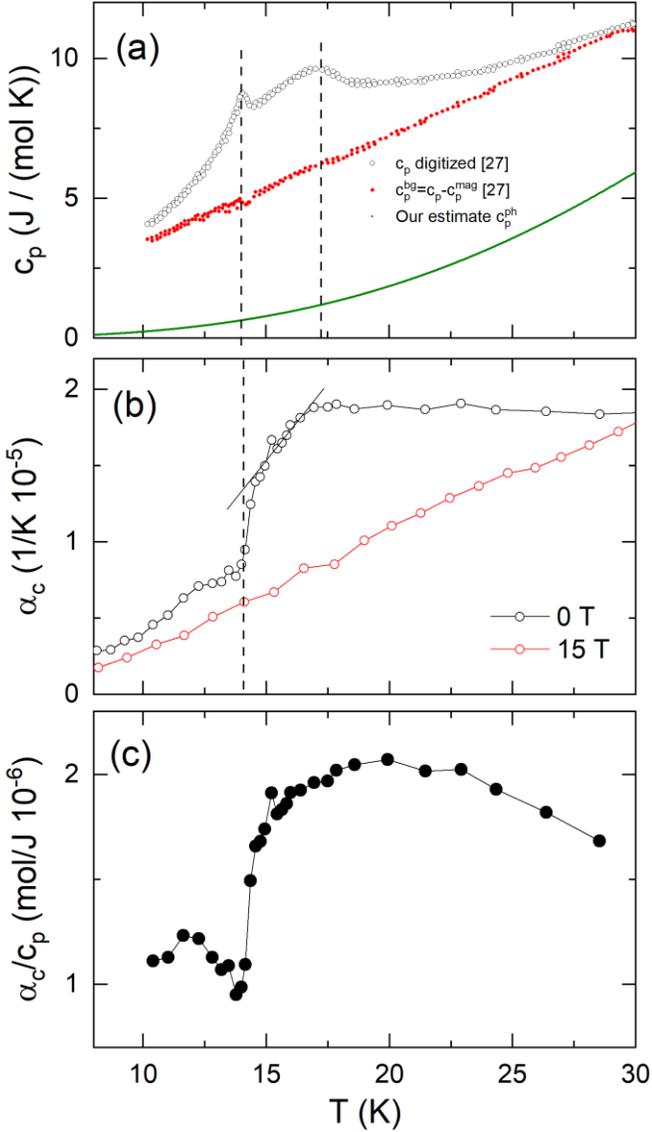

Figure. 9. (a) Specific heat $c_p$ from Ref. [27] and (b) thermal expansion coefficient $\alpha_c$ of $CrCl_3$. (c) Grüneisen ratio $c_p/\alpha_c$. In (a), we calculated the background used in [27] from the therein reported magnetic specific heat. In addition, we present our estimate of the phonon background (green line). Vertical dashed lines are guide to the eye.

According to the Ehrenfest relation, the uniaxial pressure dependence can be obtained from the experimental data by means of the anomalies $\Delta\alpha_c$ and $\Delta c_p$ in the thermal expansion coefficient and in the specific heat, respectively, at $T_N$, as following:

$$\left.\frac{\partial T_N}{\partial p_c}\right|_B = T_N V_m \frac{\Delta\alpha_c}{\Delta c_p}.$$

By using $\Delta\alpha_c = -6.1 \times 10^{-6}$ 1/K, $\Delta c_p = 0.9$ J/(mol K) and the molar volume [60] $V_m = 2.147 \times 10^{-4}$ m$^3$/mol, applying Eq. 1 yields $\partial T_N/\partial p_c = -20$ K/GPa. This corresponds to the relative pressure effect $\partial \ln T_N/\partial p_c$ of -143 %/GPa. Hence, in contrast to isostructural materials evolving ferromagnetic order, uniaxial strain applied perpendicular to the layers considerably suppresses long range magnetic (here: AFM) order in $CrCl_3$. The sign of pressure effect however agrees to the effects in α-$RuCl_3$ (see Table 2). [18,61-63]

Table 2. Experimentally determined uniaxial pressure dependencies for p||c axis of several van der Waals materials. Materials exhibiting ferromagnetic order (at $T_C$) as well as ones with antiferromagnetic order (at $T_N$) are listed. We also show the result of our Grüneisen analysis in $CrCl_3$ at the crossover temperature T*.

| | AF/FM | $T_i$ [K] | $\frac{\partial T_i}{\partial p_c}\left[\frac{K}{GPa}\right]$ | $\frac{\partial \ln T_i}{\partial p_c}\left[\frac{\%}{GPa}\right]$ | Ref. |
|---|---|---|---|---|---|
| $CrCl_3$ | $T_N$ | 14 | -20 | -143 | this work |
| $CrCl_3$ | (T*) | 17 | +7 | +43 | this work |
| $Cr_2Ge_2Te_6$ | $T_C$ | 65 | 24.7 | +38 | [61] |
| $CrI_3$ | $T_C$ | 61 | 1.7 | +3 | [18] |
| α-$RuCl_3$ | $T_{N3}$ | 7 | -10-14 | -140-200 | [62] |

Summarising the results of the Grüneisen and Ehrenfest analyses, uniaxial pressure along the c axis strongly suppresses long-range AFM order while it enhances ferromagnetic correlations above $T_N$. In an attempt to quantify the latter effect, we employ $\gamma_c(T = T^*) = 2*10^{-6}$ mol/J (Fig. 6c). By means of the Grüneisen relation this yields the uniaxial pressure dependence $\partial\epsilon/\partial p_c \approx +7$ K/GPa, at T*. While there may be several, i.e., AFM ($\epsilon_2$) and FM ($\epsilon_1$), competing energy scales relevant at T*, this result confirms the significance of ferromagnetic interactions in this temperature regime as well as their enhancement upon application of p||c. In particular, we find competing effects of uniaxial pressure similar to what is observed in the hydrostatic experiments discussed above.

### F. Pressure-dependent properties from first principles theory

We have considered two types of interlayer magnetic stacking in these calculations: ferromagnetic (FM) and antiferromagnetic (AFM) configurations, since these two were observed experimentally at zero pressure. For the FM configuration, we found that under pressure the system remains insulating and the band gap decreases only slightly by 0.1 eV between 0 and 5 GPa. For the AFM configuration, we made a similar observation where the gap decreases by 0.2 eV. Also, the Cr magnetic moment is very stable under pressure, staying equal to 3 $\mu_B$, due to the fact that the system is always insulating in the studied pressure range and we consider one-electron picture of density functional theory. We note that many-body corrections and spin-orbit coupling effects could have changed the Cr moment, but for the current study of interplay between AFM and FM configurations we do not expect such effects to play a major role. The geometry of the $CrI_3$ system was obtained from total energy (or force) minimization using DFT. For a given external pressure, P, the lattice vectors and internal atomic positions were optimized until the Hellman-Feynman forces were less than 5 meV/Å and the mechanical stress tensor was diagonal and matched the given external pressure within at least 0.1 GPa.. The enthalpy difference $\Delta H$ between the two magnetic phases, plotted in Figure 8b, indicates which phase is preferable at a given pressure: $\Delta H < 0$ implies that the AFM phase, shown in the left

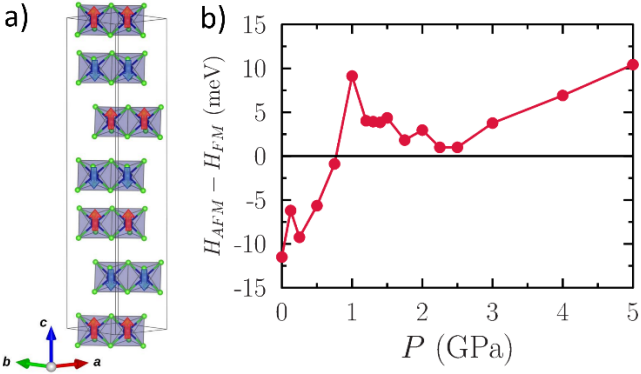

Figure 10: a) crystal structure of AFM-ordered undoped CrCl$_3$ in the hexagonal setting, b) pressure-dependence of the enthalpy difference between the AFM and FM phases of CrCl$_3$ indicating an AFM-FM transition at around 1 GPa.

part of Figure 8, is stable and $\Delta H > 0$ corresponds to a stable FM phase. experimental results of our work

It is this competition between ferromagnetic and antiferromagnetic interlayer interactions that is relevant for the The obtained DFT results, shown in Fig. 8, indicate a transition from AFM to FM interlayer stacking around 1 GPa which is consistent with experimental observations. As seen from Fig. 11, where we summarize the experimental results of the magnetism, there is a complex magnetic structure with both AFM and FM interactions at P = 0 (left side of the figure) and one component, FM, at P = 1.2 GPa (right side of the figure). Additionally, the antiferromagnetic transition shifts to higher temperatures, stabilizing the AFM order at elevated temperatures, as shown in Figure 9. At low fields (below 1,000 Oe), M(T) is characterized by a single transition, as the interlayer FM planes are not sufficiently aligned with the applied field. At higher fields (above 1,000 Oe), M(T) shows double transitions, where the interplay between ferromagnetic and antiferromagnetic order creates a transitional region. The transition temperatures, including $T_N^{onset}$ and $T_C$ at 2,000, 3,000, and 4,000 Oe, were estimated from derivative plots around the ferromagnetic regime. Initially, $T_N$ decreases with field up to 1,000 Oe and then increases up to 10,000 Oe, while the reduction in $T_C$ reflects the onset of ferromagnetic behavior.

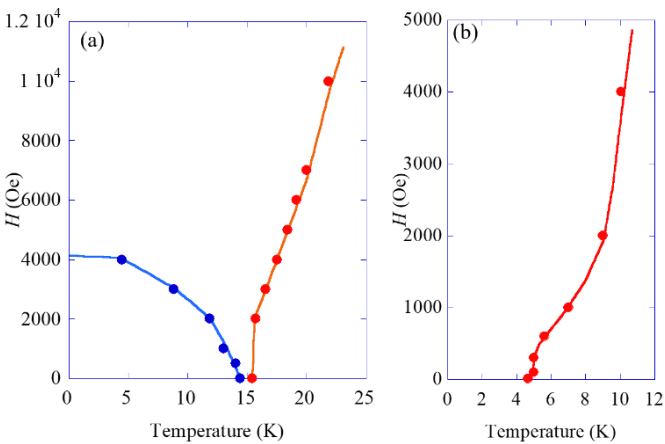

Figure 11: Néel temperature ($T_N$) and Curie temperature ($T_C$) as a function of an applied magnetic field for (a) 0 GPa and for (b) 1.2 GPa.

## IV. Conclusions

The investigation of chemically and externally applied pressure on CrCl$_3$ reveals significant pressure-induced modifications in its structural, vibrational, optical band gap, and magnetic properties. Raman measurements confirm that the smooth shift in Raman-active modes and disappearance of $A_g^3$ mode at 9.9 GPa in Fe-doped CrCl$_3$. High-pressure PL studies highlight an increase in the energy bandgap up to ~15 GPa, in contrast to the parent CrCl$_3$ sample, which exhibits a trend change at ~10 GPa. The Fe doping influences the bandgap variation in this system.. Magnetization results under pressure shows two competing magnetic components (FM and AFM) at ambient conditions, where at the lowest temperature and applied field, the FM component dominates. Significant magneto-elastic coupling is directly evidenced by our thermal expansion data which show an anomaly at $T_N$ as well as anomalous contributions and field effects at higher temperatures. In particular, Grüneisen analysis confirms the presence of competing, i.e., ferro- and antiferromagnetic interactions which dependencies on uniaxial pressure along the c axis have been determined to -143 %/GPa and +43 %/GPa, respectively. The experimental findings are in agreement with theoretical results based on DFT. In the experiments, we observe a pressure-enhanced ferromagnetic interlayer coupling that is followed by the stabilization of antiferromagnetic ordering, due to weakened direct interlayer interactions. Above 1.2 GPa the FM component of the magnetism is gone in the experimental observations, which is also in good agreement with DFT based theory. In fact, the DFT calculations result in an AFM-to-FM interlayer stacking transition at around 1 GPa. The findings reported here underscore the potential of CrCl$_3$ for use in pressure-tunable magnetic and optoelectronic applications, where, e.g., the delicate balance between FM and AFM configurations could have potential for sensor applications. In light of this, magnetoresistance experiments of the FM and AFM phases are of interest, as is the pressure induced magnetism of the Fe doped samples. Such studies are however outside the scope of the present investigation.

**ACKNOWLEDGMENT**

A.A and M.-H. acknowledges the support from the Advanced Materials Research Lab at the University of Sharjah. G.L., M.E.H, A.Abu., M.H. acknowledges the support from ARG01-0516-230179 project funded by QRDI. Support by Deutsche Forschungsgemeinschaft (DFG) under Germany's Excellence Strategy EXC2181/1-390900948 (The Heidelberg STRUCTURES Excellence Cluster) is gratefully acknowledged. L. G. acknowledges funding by the International Max-Planck Research School for Quantum Dynamics (IMPRS-QD) Heidelberg. Swedish Research Council (Vetenskapsrådet, VR) Grant No. 2016-05980, Grant No. 2019-05304, and Grant No. 2024-04986, and the Knut and Alice Wallenberg foundation Grant No. 2018.0060, Grant No. 2021.0246, and Grant No. 2022.0108 are acknowledged. The Wallenberg Initiative Materials Science for Sustainability (WISE) funded by the Knut and Alice Wallenberg Foundation is also acknowledged. The computations/data handling were enabled by resources provided by the National Academic Infrastructure for

10Supercomputing in Sweden (NAISS), partially funded by the Swedish Research Council through grant agreement no. 2022-06725. O.E. also acknowledges support from STandUPP and eSSENCE, as well as the ERC (FASTCORR- Synergy Grant No. 854843).
## References

[1] Novoselov KS, Geim AK, Morozov SV, Jiang DE, Zhang Y, Dubonos SV, Grigorieva IV, Firsov AA. Electric field effect in atomically thin carbon films. science. 2004, 306(5696), 666-9. https://doi.org/10.1126/science.1102896

[2] Lado JL, Fernández-Rossier J. On the origin of magnetic anisotropy in two dimensional CrI3. 2D Materials. 2017, 4(3), 035002. https://doi.org/10.1088/2053-1583/aa75ed

[3] Balan AP, Puthirath AB, Roy S, Costin G, Oliveira EF, Saadi MA, Sreepal V, Friedrich R, Serles P, Biswas A, Iyengar SA. Non-van der Waals quasi-2D materials; recent advances in synthesis, emergent properties and applications. Materials Today. 2022, 58, 164-200. https://doi.org/10.1016/j.mattod.2022.07.007

[4] Li C, Cao Q, Wang F, Xiao Y, Li Y, Delaunay JJ, Zhu H. Engineering graphene and TMDs based van der Waals heterostructures for photovoltaic and photoelectrochemical solar energy conversion. Chemical Society Reviews. 2018, 47(13), 4981-5037. https://doi.org/10.1039/C8CS00067K

[5] Zheng W, Liu X, Xie J, Lu G, Zhang J. Emerging van der Waals junctions based on TMDs materials for advanced gas sensors. Coordination Chemistry Reviews. 2021, 447, 214151. https://doi.org/10.1016/j.ccr.2021.214151

[6] Ma Q, Ren G, Xu K, Ou JZ. Tunable optical properties of 2D materials and their applications. Advanced Optical Materials. 2021, 9(2), 2001313. https://doi.org/10.1002/adom.202001313

[7] Xu H, Xue Y, Liu Z, Tang Q, Wang T, Gao X, Qi Y, Chen YP, Ma C, Jiang Y. Van der Waals Heterostructures for Photoelectric, Memory, and Neural Network Applications. Small Science. 2024, 4(4), 2300213. https://doi.org/10.1002/smsc.202300213

[8] Chen E, Xu W, Chen J, Warner JH. 2D layered noble metal dichalcogenides (Pt, Pd, Se, S) for electronics and energy applications. Materials Today Advances. 2020, 7, 100076. https://doi.org/10.1016/j.mtadv.2020.100076

[9] McGuire MA. Cleavable magnetic materials from van der Waals layered transition metal halides and chalcogenides. Journal of Applied Physics. 2020, 128(11). https://doi.org/10.1063/5.0023729

[10] Castellanos-Gomez A, Duan X, Fei Z, Gutierrez HR, Huang Y, Huang X, Quereda J, Qian Q, Sutter E, Sutter P. Van der Waals heterostructures. Nature Reviews Methods Primers. 2022, 2(1), 58. https://doi.org/10.1038/s43586-022-00139-1

[11] Elahi E, Dastgeer G, Nazir G, Nisar S, Bashir M, Qureshi HA, Kim DK, Aziz J, Aslam M, Hussain K, Assiri MA. A review on two-dimensional (2D) magnetic materials and their potential applications in spintronics and spin-caloritronic. Computational Materials Science. 2022, 213, 111670. https://doi.org/10.1016/j.commatsci.2022.111670

[12] Wang QH, Bedoya-Pinto A, Blei M, Dismukes AH, Hamo A, Jenkins S, Koperski M, Liu Y, Sun QC, Telford EJ, Kim HH. The magnetic genome of two-dimensional van der Waals materials. ACS nano. 2022, 16(5), 6960-7079. https://doi.org/10.1021/acsnano.1c09150

[13] Huang B, Clark G, Navarro-Moratalla E, Klein DR, Cheng R, Seyler KL, Zhong D, Schmidgall E, McGuire MA, Cobden DH, Yao W. Layer-dependent ferromagnetism in a van der Waals crystal down to the monolayer limit. Nature. 2017, 546(7657), 270-3. https://doi.org/10.1038/nature22391

[14] Thiel L, Wang Z, Tschudin MA, Rohner D, Gutiérrez-Lezama I, Ubrig N, Gibertini M, Giannini E, Morpurgo AF, Maletinsky P. Probing magnetism in 2D materials at the nanoscale with single-spin microscopy. Science. 2019, 364(6444), 973-6. https://doi.org/10.1126/science.aav6926

[15] Basak K, Ghosh M, Chowdhury S, Jana D. Theoretical studies on electronic, magnetic and optical properties of two dimensional transition metal trihalides. Journal of Physics: Condensed Matter. 2023, 35(23), 233001. https://doi.org/10.1088/1361-648X/acbffb

[16] Mermin ND, Wagner H. Absence of ferromagnetism or antiferromagnetism in one-or two-dimensional isotropic Heisenberg models. Physical Review Letters. 1966, 17(22), 1133. https://doi.org/10.1103/PhysRevLett.17.1133

[17] Yang Q, Hu X, Shen X, Krasheninnikov AV, Chen Z, Sun L. Enhancing ferromagnetism and tuning electronic properties of CrI3 monolayers by adsorption of transition-metal atoms. ACS Applied Materials & Interfaces. 2021, 13(18), 21593-601. https://doi.org/10.1021/acsami.1c01701

[18] J. Arneth, M. Jonak, S. Spachmann, M. Abdel-Hafiez, Y. O. Kvashnin, and R. Klingeler, Uniaxial pressure effects in the two-dimensional van der Waals ferromagnet CrI3, Phys. Rev. B 105, L060404 (2022). https://doi.org/10.1103/PhysRevB.105.L060404

[19] Anirudha Ghosh, Deobrat Singh, T Aramaki, Qingge Mu, Vladislav Borisov, Yaroslav Kvashnin, G Haider, M Jonak, D Chareev, SA Medvedev, R Klingeler, M Mito, EH Abdul-Hafidh, J Vejpravova, M Kalbàč, Rajeev Ahuja, Olle Eriksson, Mahmoud Abdel-Hafiez, Exotic magnetic and electronic properties of layered single crystals under high pressure. Phys. Rev. B 105, L081104 (2022) https://doi.org/10.1103/PhysRevB.105.L081104

[20] M Jonak, E Walendy, J Arneth, Mahmoud Abdel-Hafiez, R Klingeler, Low-energy magnon excitations and emerging anisotropic nature of short-range order in CrI3, Phys. Rev. B 106, 214412 (2022). https://doi.org/10.1103/PhysRevB.106.214412

[21] McGuire MA, Dixit H, Cooper VR, Sales BC. Coupling of crystal structure and magnetism in the layered, ferromagnetic insulator CrI3. Chemistry of Materials. 2015, 27(2), 612-20. https://doi.org/10.1021/cm504242t

[22] Liu Y, Wu L, Tong X, Li J, Tao J, Zhu Y, Petrovic C. Thickness-dependent magnetic order in CrI3 single crystals. Scientific Reports. 2019, 9(1), 13599. https://doi.org/10.1038/s41598-019-50000-x